\documentstyle[prd,preprint,tighten,aps,
amssymb,newlfont,bbm]{revtex}
\begin{document}
\preprint{
\begin{tabular}{r}
UWThPh-2002-14\\
October 2002
\end{tabular}
}
\draft
\title{
On the Orthogonality of Independently Propagating
States as Occurring in the Lee--Oehme--Yang Theory
}
\author{G.V. Dass}
\address{Physics Department, Indian Institute of Technology \\
\small Powai, Bombay 400076, India}
\author{W. Grimus}
\address{Institut f\"ur Theoretische Physik, Universit\"at Wien \\
\small Boltzmanngasse 5, A--1090 Wien, Austria}

\maketitle

\begin{abstract}
We generalize a theorem by Khalfin, originally 
derived for the states 
$| F_1 \rangle = | M^0 \rangle$, 
$| F_2 \rangle = |\bar M^0 \rangle$, where $M^0$ is a neutral
flavoured meson (e.g., $K^0$ or $B_d^0$), by assuming CPT invariance.
Dispensing with CPT invariance and allowing for an arbitrary pair
of orthogonal states $| F_{1,2} \rangle$, 
we show that any linear combinations
$| P_a \rangle = p_a | F_1 \rangle + q_a | F_2 \rangle$ and
$| P_b \rangle = p_b | F_1 \rangle - q_b | F_2 \rangle$,
if postulated to be independently
propagating in time, as in the Lee--Oehme--Yang Theory, must 
be mutually orthogonal.
This implies a reciprocity relation: equality of
the probabilities of the transitions 
$| F_1 \rangle \leftrightarrow | F_2 \rangle$. 
Also implied is another
relation involving the coefficients $p_{a,b}$, $q_{a,b}$,
which can be interpreted as 
$\mathrm{Im}\, \theta = 0$, where $\theta$ is the rephasing-invariant
parameter describing CPT violation in $M^0 \bar M^0$ mixing for
Khalfin's choice of $| F_{1,2} \rangle$.
The states $| F_{1,2} \rangle$ of \emph{our} theorem
need not form a particle-antiparticle
pair, nor even be restricted to particle physics.
\end{abstract}


\newpage

The complex formed by a neutral meson $M^0$ (i.e., $K^0$, $D^0$,
$B_d^0$, $B_s^0$) and its antiparticle $\bar M^0$ is a very important
testing ground for discrete symmetries and physics beyond the Standard
Model. Its theoretical treatment relies heavily on the
Weisskopf--Wigner approximation \cite{wwa} and the
Lee--Oehme--Yang (LOY) Theory \cite{loy} (for reviews see, for instance,
Refs.~\cite{kabir-book,luis-book,bigi-book}). Central to this approximate
theory are the linear combinations
\begin{equation}\label{MHML}
\begin{array}{cc}
| M_H \rangle = p_H | M^0 \rangle + q_H | \bar M^0 \rangle \,,
& |p_H|^2 + |q_H|^2 = 1 \,, \\
| M_L \rangle = p_L | M^0 \rangle - q_L | \bar M^0 \rangle \,,
& |p_L|^2 + |q_L|^2 = 1 \,, 
\end{array}
\end{equation}
with complex constants $p_{H,L}$ and $q_{H,L}$, which propagate
independently in time. This property is also called ``lack of vacuum
regeneration''. The subscripts $H$ and $L$ refer to heavy and
light, respectively. By starting with the additional assumption of
\begin{equation}\label{cpt-inv}
\mbox{CPT invariance in $M^0 \bar M^0$ mixing:} \quad
\frac{q_H}{p_H} = \frac{q_L}{p_L} \,,
\end{equation}
Khalfin demonstrated \cite{khalfin97,khalfin-report}
that the independent propagation of 
$| M_H \rangle$, $| M_L \rangle$ and positivity of the spectrum of the
total Hamiltonian governing the time
evolution of $| M^0 \rangle$, $| \bar M^0 \rangle$ and of the decay
states $| f \rangle$ 
leads to $\langle M_H | M_L \rangle = 0$ (``Main Theorem,
Part (A)'' in Ref.~\cite{khalfin97}).
This result is apparently at odds with the experimental finding of CP
violation in $K^0 \bar K^0$ mixing \cite{cp,cp1}, which is equivalent to
$\langle K_L | K_S \rangle \neq 0$, if CPT invariance holds 
in $K^0 \bar K^0$ mixing. 
Note that in the kaon system we make the identification 
$| K_L \rangle = | M_H \rangle$, $| K_S \rangle = | M_L \rangle$.
Khalfin's purpose was to demonstrate forcefully, on the basis of
general principles, the inadequacy of the LOY Theory as a good theory
of CP violation. While the mathematical content of his
theorem remains true, his claim of a resulting large
``new CP violation effect''
was subsequently refuted in Refs.~\cite{chiu,nowakowski}; 
we shall return to this point later.

In this paper, we generalize Khalfin's Theorem by dropping the
requirement of CPT invariance.\footnote{In an unpublished report
\cite{khalfin94}, quoted as Ref.~[30] in his paper \cite{khalfin97}, 
Khalfin relaxed the assumption of CPT invariance.} 
It is interesting to relax this requirement 
in view of the need of improvement of present data for many of the
tests of CPT invariance \cite{RPP02}. Because of our generalization, 
we go beyond the choice of $| F_{1,2} \rangle$ being a
particle-antiparticle pair, though, at present, 
this choice seems to be the only one with an experimental application
in particle physics.\footnote{Thus our theorem 
is not restricted to particle physics. See remarks at the end of
Ref.~\cite{chiu} for its applicability in the study of 
communicating metastable states in atomic physics.}

We consider a pair of orthonormal electrically
neutral states $| F_{1,2} \rangle$, in analogy to $| M^0 \rangle$ and
$|\bar M^0 \rangle$, and states $| f \rangle$ orthogonal to 
$| F_{1,2} \rangle$. Then the time evolution of the states 
$| F_{1,2} \rangle$ is, most generally, expressed as
\begin{equation}\label{timeev}
| F_j(t) \rangle = \sum_{k=1,2} | F_k \rangle A_{kj}(t) + 
\sum_f c_{fj}(t) | f \rangle 
\quad \mathrm{with} \quad 
A_{kj}(t) = \langle F_k | e^{-iHt} | F_j \rangle 
\quad \forall\, t \geq 0.
\end{equation}
Now we want to be more specific about the time evolution
(\ref{timeev}). We postulate the existence of
two linearly independent superpositions $| P_{a,b} \rangle$ of
$| F_{1,2} \rangle$, defined by
\begin{equation}\label{PP}
\begin{array}{cc}
| P_a \rangle = p_a | F_1 \rangle + q_a | F_2 \rangle \,, &
| p_a |^2 + | q_a |^2 = 1 \,, \\
| P_b \rangle = p_b | F_1 \rangle - q_b | F_2 \rangle \,, &
| p_b |^2 + | q_b |^2 = 1 \,,
\end{array}
\end{equation}
where $p_{a,b}$ and $q_{a,b}$ are complex
constants, such
that these states propagate independently in time. 
Eq.~(\ref{PP}) is the analogue of Eq.~(\ref{MHML}). 
This means that
the time evolution of $| P_{a,b} \rangle$ is described for $t \geq 0$ by
\begin{equation}\label{LVR}
\begin{array}{ccc}
| P_a \rangle & \stackrel{t}{\to} &
\lambda_a(t) | P_a \rangle + \sum_f c_{fa}(t) | f \rangle \,, \\
| P_b \rangle & \stackrel{t}{\to} &
\lambda_b(t) | P_b \rangle + \sum_f c_{fb}(t) | f \rangle \,,
\end{array}
\end{equation}
where $\lambda _{a,b}$ are suitable propagation
functions and the $| f \rangle$ terms in this equation are directly
related to those in Eq.~(\ref{timeev}). The main point in Eq.~(\ref{LVR}) 
is that there are no transitions
$| P_a \rangle \stackrel{t}{\to} | P_b \rangle$ or
$| P_b \rangle \stackrel{t}{\to} | P_a \rangle$.

Now we formulate our theorem which generalizes the ``Main Theorem,
Part (A)'' of Khalfin in Ref.~\cite{khalfin97}.

\vspace{1.5mm}

\noindent
\textbf{Theorem:}
\begin{itshape}
If the spectrum of the Hamiltonian $H$ is positive (or, more
generally, bounded from below), then the two independently propagating
states $| P_{a,b} \rangle$ are mutually orthogonal, i.e.,
$\langle P_a | P_b \rangle = 0$.
\end{itshape}

\vspace{1.5mm}

\noindent
\textbf{Proof:}
The time evolution as expressed by Eq.~(\ref{timeev}) is connected
with Eq.~(\ref{LVR}) via
\begin{equation}
\mathrm{diag}\,(\lambda_a(t), \lambda_b(t)) \equiv B(t) =
R^{-1} A(t) R 
\quad \mathrm{with} \quad
R = \left(
\begin{array}{lr} p_a & p_b \\ q_a & -q_b \end{array}
\right).
\end{equation}
Now, the independent propagation of the states $| P_{a,b} \rangle$ is
formulated as \cite{dass}
\begin{equation}
\begin{array}{c}
B_{21} =
\left\{ (A_{11} - A_{22}) p_a q_a + A_{12} q_a^2 - A_{21} p_a^2
\right\}/D = 0 \,, \\
B_{12} =  
\left\{ (A_{11} - A_{22}) p_b q_b - A_{12} q_b^2 + A_{21} p_b^2
\right\}/D = 0 \,,
\end{array}
\end{equation}
where $D = p_a q_b + p_b q_a = -\det R$.
From these equations we derive \cite{dass}
\begin{eqnarray}
A_{12} & = & \frac{p_a p_b}{q_a q_b}\, A_{21} \,, \label{1221} \\
A_{11} - A_{22} & = & 
\left( \frac{p_a}{q_a} - \frac{p_b}{q_b} \right) A_{21} \,.
\label{1122}
\end{eqnarray}

Following Khalfin \cite{khalfin97} (see also Refs.~\cite{chiu,kabir95}), 
we introduce the expansions
\begin{equation}
| F_j \rangle = \int_0^\infty \mathrm{d}m\, \sum_\alpha d_{j\alpha}(m)\,
| \phi _\alpha (m) \rangle \quad (j=1,2)
\end{equation}
over the complete set of orthonormal
eigenfunctions $| \phi _\alpha (m) \rangle$ 
of the full Hamiltonian $H$:
\begin{equation}
H\, | \phi _\alpha (m) \rangle = m\, | \phi_\alpha (m) \rangle
\quad \mathrm{with} \quad
\langle \phi_\alpha (m) | \phi_\beta (m') \rangle = 
\delta _{\alpha \beta }\, \delta (m - m') \,;
\end{equation}
the $m$-integration is over the non-negative
spectrum of $H$. The matrix elements $A_{kj}(t)$ of the time
evolution (\ref{timeev}) are expressed as
\begin{equation}
A_{kj}(t) = \int_0^\infty \mathrm{d}m\, e^{-imt} C_{kj}(m)
\quad \mathrm{with} \quad
C_{kj}(m) = \sum_\alpha d_{k\alpha}^*(m) d_{j\alpha}(m) \,.
\end{equation}
Then, Eq.~(\ref{1221}) means
\begin{equation}\label{12}
\int_0^\infty \mathrm{d}m\, e^{-imt} 
\left[ q_a q_b\, C_{12}(m) - p_a p_b\, C_{21}(m) \right] = 0 
\quad \forall\, t \geq 0 \,.
\end{equation}
We would now like to infer from this equation that the integrand
itself is zero. To make this conclusion we need to know that the
integral is zero also for negative times $t$. 
In Refs.~\cite{khalfin97,chiu,kabir95} an analyticity argument is used
to justify this conclusion; however, the details of this argument are 
not worked out there. In the appendix we present a full proof that from
Eq.~(\ref{12}) it follows indeed that the integral is zero for all
times $t$ and that, consequently, the integrand is zero as 
well.\footnote{Note that it would be reasonable to assume that the 
integrand---as a product of
square-integrable functions---is absolutely integrable. 
However, because of the mathematical intricacies involved in the proof
presented in the appendix, we
seem to be forced to assume that the integrand in Eq.~(\ref{12}) is
square-integrable.} 

If the function $C_{12}(m) = C_{21}^*(m)$ is identically zero, 
the mutually orthogonal states 
$| F_{1,2} \rangle$ themselves propagate independently in time,
because in this case we have $A_{12}(t) = A_{21}(t) = 0$ 
for all $t \geq 0$. Let us assume now that $C_{12}(m)$ is not
identically zero. We choose a value $m_0$ such that 
$C_{12}(m_0) \neq 0$. Then from the vanishing of the integrand in
Eq.~(\ref{12}) we obtain
\begin{equation}\label{cond1}
\frac{p_a p_b}{q_a q_b} = \frac{C_{12}(m_0)}{C_{21}(m_0)} \,.
\end{equation}
Along the same lines, we derive from Eq.~(\ref{1122}) that
\begin{equation}\label{cond2}
C_{11}(m_0) - C_{22}(m_0) = 
\left( \frac{p_a}{q_a} - \frac{p_b}{q_b} \right) C_{21}(m_0) \,.
\end{equation}
It is useful to define the quantities
\begin{equation}\label{def}
\rho_a = \frac{p_a}{q_a} e^{-i\gamma} \,, \;
\rho_b = \frac{p_b}{q_b} e^{-i\gamma}
\quad \mathrm{with} \quad
C_{12}(m_0) = |C_{12}(m_0)| e^{i\gamma} \,.
\end{equation}
Note that $\rho_{a,b}$ are invariant under the rephasings 
$| F_j \rangle \to | F_j' \rangle \equiv e^{i\delta_j} | F_j \rangle$ of
the states we started with (for a thorough discussion of rephasings in
the $M^0 \bar M^0$ system, see Ref.~\cite{luis-book}). Then, 
Eq.~(\ref{cond1}) can be rewritten as 
\begin{equation}\label{condI}
\rho_a \rho_b = 1 \,.
\end{equation}
The analogous rewriting of Eq.~(\ref{cond2}) yields
\begin{equation}
\rho_a - \rho_b = \frac{C_{11}(m_0) - C_{22}(m_0)}{|C_{12}(m_0)|} \,.
\end{equation}
The right-hand side of this equation is real but its value is
unknown. Therefore, it is not useful for constraining
$\rho_{a,b}$. However, its imaginary part gives
\begin{equation}\label{condII}
\mathrm{Im}\, \rho_a = \mathrm{Im}\, \rho_b \,.
\end{equation}
A little algebra shows that Eqs.~(\ref{condI}) and (\ref{condII}) 
together imply
\begin{equation}\label{im}
\mathrm{Im}\, \rho_a = \mathrm{Im}\, \rho_b = 0 \,.
\end{equation}
In other words, we finally have
\begin{equation}\label{result}
\rho_{a,b} \in \mathbbm{R} 
\quad \mathrm{and} \quad
\rho_a \rho_b = 1 \,.
\end{equation}

Now we consider the overlap between $| P_a \rangle$ and $| P_b
\rangle$:
\begin{equation}
\langle P_a | P_b \rangle = p_a^* p_b - q_a^* q_b = 
q_a^* q_b\, ( \rho_a^* \rho_b - 1) \,.
\end{equation}
Using Eq.~(\ref{result}), we find $\langle P_a | P_b \rangle = 0$,
Q.E.D.

We mention two corollaries of our theorem.
Firstly, Eq.~(\ref{cond1}), when used in Eq.~(\ref{1221}), directly
gives the reciprocity relation \cite{khalfin97,khalfin-report}
\begin{equation}\label{reciprocity}
|A_{12}(t)| = |A_{21}(t)| 
\end{equation}
for the transitions $| F_1 \rangle \leftrightarrow | F_2 \rangle$;
this T-invariance relation is general, no assumptions about the nature of the
states $| F_{1,2} \rangle$ or about symmetries have been 
made.\footnote{That the relation (\ref{reciprocity})
does not require CPT invariance was already pointed out in
Ref.~\cite{kabir95}.} For the 
$M^0 \bar M^0$ system, this T invariance in mixing
\cite{aharony} is usually stated in the form \cite{luis-book}
\begin{equation}\label{Tinv}
\left| \frac{q}{p} \right| = 1 \quad \mathrm{with} \quad
\frac{q}{p} = \sqrt{\frac{q_a q_b}{p_a p_b}} \,.
\end{equation}
For the second corollary, we define the quantity
\begin{equation}
\theta = \frac{(q_a/p_a) - (q_b/p_b)}{(q_a/p_a) + (q_b/p_b)} =
\frac{\rho_b - \rho_a}{\rho_b + \rho_a} \,.
\end{equation}
Then, Eq.~(\ref{im}) gives
\begin{equation}\label{Imtheta}
\mathrm{Im}\, \theta = 0 \,.
\end{equation}
Eqs.~(\ref{Tinv}) and (\ref{Imtheta}) are already known
\cite{luis-book,luis} to be equivalent to the
vanishing of $\langle P_a | P_b \rangle$ for the particular choice
$| F_1 \rangle = | M^0 \rangle$, 
$| F_2 \rangle = | \bar M^0 \rangle$. Our derivation is more general;
the $| F_{1,2} \rangle$ need not form a particle-antiparticle 
pair.\footnote{Note that $\langle P_a | P_b \rangle = 0$
means that the effective Hamiltonian matrix in the subspace generated by 
$| F_{1,2} \rangle$ is a normal matrix \cite{luis-book}.}
We want to stress that, if we specialize to the $M^0 \bar M^0$ system, 
full CPT invariance in mixing 
(viz.\ $\theta = 0$ \cite{luis-book,aharony,kabir70}) 
is not required by the theorem; just Eq.~(\ref{Imtheta}) is required.

Concentrating on the $K^0 \bar K^0$ system, we want to compare the
requirement $\langle K_L | K_S \rangle = 0$ of our theorem with
available experimental data.
Using data on various kaon decay channels and allowing for violations
of the discrete symmetries CPT, CP 
and T and of the $\Delta S = \Delta Q$ rule, the evaluation of the
Bell--Steinberger unitarity relation by the CPLEAR Collaboration
\cite{cp1} gives the result\footnote{Note that the quantity 
$1 - |q/p|$ is called $2\, \mathrm{Re}\, \varepsilon$ 
in the CPLEAR paper; see Ref.~\cite{dass02} for a comparison with the
CPLEAR notation.}  
$1 - |q/p| = (3.30 \pm 0.05) \times 10^{-3}$. This result\footnote{The
same result, with a much less compelling accuracy
$1 - |q/p| = (3.1 \pm 0.7 \pm 0.5) \times 10^{-3}$, was recently
obtained by a ``direct observation'' of T non-invariance by the 
CPLEAR Collaboration \cite{cp,cplear}; however, 
this result for $|q/p|$ assumed CPT invariance for the
amplitudes of semileptonic neutral kaon decays. A completely
assumption-independent ``direct measurement'' of $|q/p|$ is not available at
present; see Ref.~\cite{dass01} and references cited therein.} 
continues to
be far from zero even if the assumption of the saturation of the
unitarity relation by known channels is relaxed within the limits
allowed by the experimental errors on the decay rates. 
The Bell--Steinberger unitarity relation is a consequence of the LOY
Theory and the Weisskopf--Wigner approximation. Thus, the CPLEAR
result of non-zero $1 - |q/p|$, which leads to a non-vanishing 
overlap $\langle K_L | K_S \rangle$, is in contradiction to Khalfin's
Theorem and its extension proven in the present paper.
On the other hand, 
$\mathrm{Re}\, \theta$ and $\mathrm{Im}\, \theta$ are both consistent
with zero \cite{cp1,cplear}.

Though outside the sope of the paper, one might ask the question 
``What goes wrong when the 
mathematically correct theorem of Khalfin is applied to the $M^0 \bar M^0$
system?'' The answer was clearly given in Refs.~\cite{chiu,nowakowski,kabir95}:
The independent propagation of $| M_H \rangle$ and $| M_L \rangle$ is not
strictly true for all times. There is indeed vacuum regeneration to a certain
degree and one might nourish the hope that future experiments provide
some clue to the size of this effect. However, the ``new
CP violation effect'', predicted by Khalfin 
\cite{khalfin97,khalfin-report} on the basis of this vacuum
regeneration, was an overestimate by many orders of 
magnitude \cite{chiu,nowakowski}.

In summary, starting with two suitable 
states $| F_{1,2} \rangle$ which mix to form two independently
propagating states $| P_{a,b} \rangle$ as occurring in the
LOY Theory, we have shown
the vanishing of the overlap of $| P_{a,b} \rangle$,
i.e., $\langle P_a | P_b \rangle = 0$. We also find the 
reciprocal transitions 
$| F_1 \rangle \leftrightarrow | F_2 \rangle$ to be equally strong,
viz.\ T invariance in mixing.
The restriction to $| F_{1,2} \rangle$ forming a
particle-antiparticle pair is not required, nor
the assumption of CPT invariance in the particle-antiparticle 
case. However, if $| F_{1,2} \rangle$ do form a 
particle-antiparticle pair, partial CPT invariance is also required:
The imaginary part of the CPT-violating parameter $\theta$ must be
zero, whereas the real part of $\theta$ is allowed by the theorem to
be non-zero.

\acknowledgments
We thank P.K. Kabir and L. Lavoura for
important communications.

\setcounter{equation}{0}
\renewcommand{\theequation}{A\arabic{equation}}
\section*{Appendix}

We now present a proof that from Eq.~(\ref{12}) it follows that the function
\begin{equation}
f(m) \equiv q_a q_b\, C_{12}(m) - p_a p_b\, C_{21}(m)
\end{equation}
must be zero.

\vspace{2mm}

\noindent
\textbf{Lemma:} 
\begin{itshape}
Let $f$ be a square-integrable function over the real numbers, which is
zero on the negative real axis. Defining a function $\hat f$ over all real
numbers by
$$
\hat f(t) = \int_0^\infty \mathrm{d}m\: e^{-itm} f(m) 
$$
and assuming that $\hat f(t) = 0$ $\forall\, t \geq 0$, it follows that 
$\hat f(t) = 0$ for negative $t$ as well. Consequently, the function $f$
itself is zero.
\end{itshape}

\vspace{2mm}

\noindent
\textbf{Proof:} The function $\hat f(t)$ defined for all real $t$ can be
extended to complex numbers $z = t -i\eta$ ($\eta > 0$) in the lower complex
half plane, where $\hat f(z)$ is an analytic function. Cauchy's Theorem allows
us to write
\begin{equation}\label{cauchy}
-\frac{1}{2\pi i} \int_{-\infty}^\infty \mathrm{d}u \, 
\frac{\hat f(u)}{u-z} = \hat f(z) \equiv
\int_0^\infty \mathrm{d}m\: e^{-izm} f(m) \,.
\end{equation}
Since $f$ is square-integrable, i.e., $f \in L^2(\mathbbm{R})$, the same holds
for $\hat f$. Actually, since $f$ is zero on the negative real axis, this
function belongs to the class of Hardy functions, for which relation
(\ref{cauchy}) is common usage (see, for instance, Ref.~\cite{dym}). Note that
the integral on the left-hand side of Eq.~(\ref{cauchy}) is well-defined
since its integrand---as a product of $L^2$ functions---is absolutely
integrable, i.e., an element of $L^1(\mathbbm{R})$.
Now we assume that $\mathrm{Re}\, z = t > 0$ and take the limit 
$\mathrm{Im}\, z = -\eta \uparrow 0$. Then the right-hand side of
Eq.~(\ref{cauchy}) becomes 0 and we arrive at
\begin{equation}\label{integral}
\int_0^\infty \mathrm{d}u\, \frac{g(u)}{u+t} = 0 \quad \forall \, 
t > 0 \,,
\end{equation}
where we have defined $g(u) = \hat f(-u)$ $\forall\, u \geq 0$. 
Furthermore, from this equation we derive
\begin{equation}\label{integral1}
\frac{1}{t-1} \left( \int_0^\infty \mathrm{d}u\, \frac{g(u)}{u+1} -
\int_0^\infty \mathrm{d}u\, \frac{g(u)}{u+t} \right) =
\int_0^\infty \mathrm{d}u\, \frac{g(u) }{(u+1)(u+t)} = 0 \quad
\forall\, t \neq 1 \,.
\end{equation}
Because of the uniform convergence of the integral (\ref{integral1}), we are
allowed to drop the requirement $t \neq 1$ and the second relation in
Eq.~(\ref{integral1}) holds thus for all positive $t$.
Now we define $G(u) = g(u)/(u+1)$ for $u \geq 0$, which is an
absolutely integrable function in $u$. Then, 
Eq.~(\ref{integral1}) is reformulated as
\begin{equation}\label{laplace}
\int_0^\infty \mathrm{d}u\, G(u) \int_0^\infty \mathrm{d}s\, e^{-s(u+t)} =
\int_0^\infty \mathrm{d}s\, e^{-ts} \int_0^\infty \mathrm{d}u\, e^{-su} \, 
G(u) = \left( \mathcal{L} \mathcal{L} G \right) (t) = 0 \,,
\end{equation}
where $\mathcal{L}$ denotes the Laplace transform with integration interval
from zero to infinity.\footnote{One of us
(W.G.) is grateful to H. Neufeld for suggesting the use of the Laplace
transform.} It is easy to argue by using the Theorem of Fubini that, because
$G$ is absolutely integrable in $u$, the exchange of the succession of
integrations in Eq.~(\ref{laplace}) is a correct operation. 
If the Laplace transform of
a function is zero, then the function itself is zero (see, for
instance, Ref.~\cite{marsden}). 
Thus we have proved that $G$ is zero, therefore, also $\hat f = 0$ and,
finally, we find $f = 0$, Q.E.D.

\end{document}